\pdfoutput=1

\documentclass[conference]{IEEEtran}
\IEEEoverridecommandlockouts
\usepackage{graphicx}
\usepackage[utf8]{inputenc}
\usepackage{times}
\fontfamily{ptm}\selectfont
\usepackage{tabularx}
\usepackage{ragged2e}
\usepackage[singlelinecheck=false]{caption}
\usepackage[T1]{fontenc}
\usepackage[numbers]{natbib}
\usepackage{amsmath}
\usepackage{amssymb}
\usepackage{xcolor}
\usepackage{mathtools}

\usepackage{babel}
\usepackage{caption}
\usepackage{subcaption}

\begin{document}

\title{Exploring ISAC Technology for UAV SAR Imaging}
\author{\IEEEauthorblockN{
Stefano Moro,
Francesco Linsalata, Marco Manzoni,
Maurizio Magarini and
Stefano Tebaldini}
\IEEEauthorblockA{
\small Department of Electronics, Information and Bioengineering, Politecnico di Milano, \textit{Italy}}
\thanks{.}
}

\maketitle

\begin{abstract}
This paper illustrates the potential of an Integrated Sensing and Communication (ISAC) system, operating in the sub-6 GHz frequency range, for Synthetic Aperture Radar (SAR) imaging via an Unmanned Aerial Vehicle (UAV) employed as an aerial base station. The primary aim is to validate the system's ability to generate SAR imagery within the confines of modern communication standards, including considerations like power limits, carrier frequency, bandwidth, and other relevant parameters. The paper presents two methods for processing the signal reflected by the scene. Additionally, we analyze two key performance indicators for their respective fields, the Noise Equivalent Sigma Zero (NESZ) and the Bit Error Rate (BER), using the  QUAsi Deterministic RadIo channel GenerAtor (QuaDRiGa), demonstrating the system's capability to image buried targets in challenging scenarios. The paper shows simulated Impulse Response Functions (IRF) as possible pulse compression techniques under different assumptions.  An experimental campaign is conducted to validate the proposed setup by producing a SAR image of the environment captured using a UAV flying with a Software-Defined Radio (SDR) as a payload. 
\end{abstract}

\section{Introduction}
\label{sec:Introduction}
In recent times, a notable surge in technological progress has greatly expanded the capabilities of Unmanned Aerial Vehicles (UAVs) across various applications~\cite{meng_uav-enabled_2023}. These advancements include enhancements in battery life, augmented reliability, the development of advanced software for autonomous navigation, and the utilization of lightweight materials~\cite{petritoli_reliability_2017}. In addition, the reduced cost of drones has substantially improved the feasibility of deploying UAV networks~\cite{zeng_wireless_2016} and presented possible solutions for continuous operations by coordinating fly and recharge operations~\cite{trotta2017fly}.

Search and rescue operations are paramount among the many compelling applications of UAVs. Their rapid deployment, ease of operation, and adaptability make them indispensable in saving lives during emergencies~\cite{cui_integrating_2021}. Equipping drones with various sensors, including optical and thermal cameras, LiDARs, and radars, prepares them for rescue missions of different natures. Notably, radar stands out for its ability to send Electromagnetic (EM) waves into the ground, enabling the detection of victims buried under debris or snow, a critical capability after events like earthquakes or avalanches~\cite{grathwohl_taking_2022}. Moreover, this same capacity can also provide essential wireless connectivity to rescuers, often unavailable in disaster-affected or remote areas~\cite{linsalataUAV}.

The Integrated Sensing and Communication (ISAC) paradigm offers a seamless solution for both problems. Here, radar sensing operations can share the spectrum or even the same signal as the communication link. The literature features numerous studies delving into various ISAC applications. For example, in~\cite{cui_integrating_2021}, several potential ISAC applications are explored from the perspective of Content Service Providers (CSPs) in scenarios like Smart Home, Sensing as a Service, Environmental monitoring, and more. The work in~\cite{pin_tan_integrated_2021} delves into UAV applications and optimizes waveform design for ISAC, while~\cite{zeng_wireless_2016} provides a comprehensive account of three UAV applications within a communications network. Moreover, in~\cite{meng_uav-enabled_2023}, it details a thorough review of single and multi-UAV ISAC, examining the pros and cons of drone swarms.
\begin{figure*}[!htbp]
     \centering
     \begin{subfigure}[b]{\columnwidth}
         \centering
         \includegraphics[width=0.7\columnwidth]{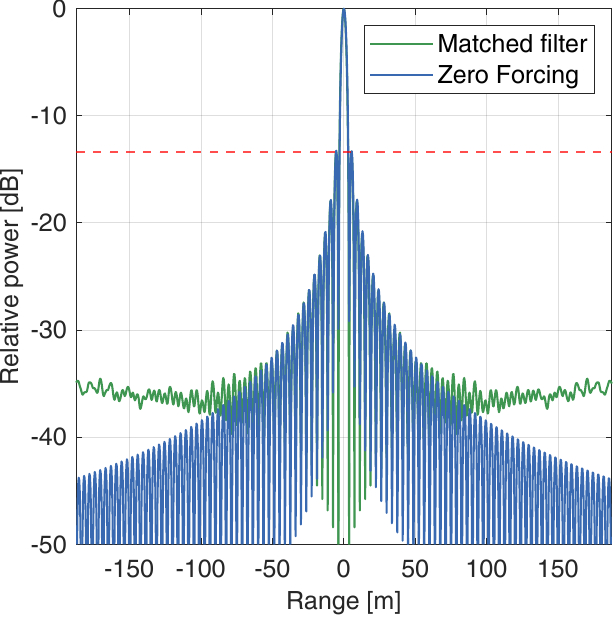}
         \caption{Impulse response with a QPSK modulated sub-carriers.}
         \label{fig:QPSK_corr}
     \end{subfigure}
\hfill
     \begin{subfigure}[b]{\columnwidth}
         \centering
         \includegraphics[width=0.7\columnwidth]{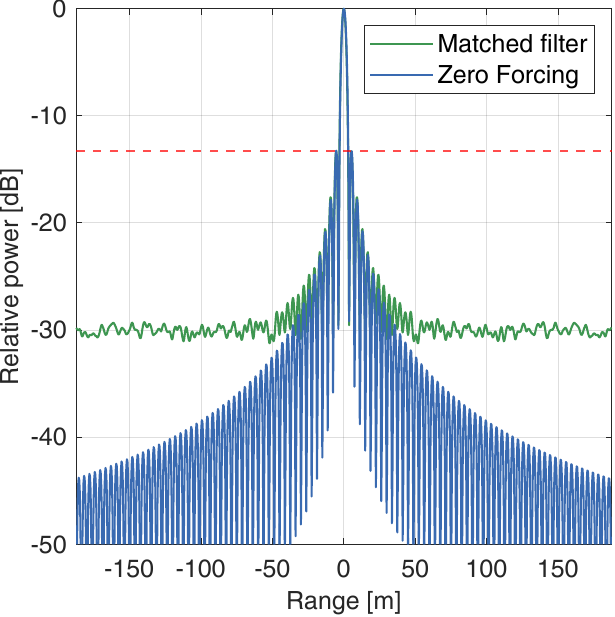}
         \caption{Impulse response with a 256-QAM modulated sub-carriers.}
         \label{fig:QAM_corr}
     \end{subfigure}
     \caption{Range compression of one OFDM symbol with QPSK and 256-QAM modulated sub-carriers. The red-dotted line highlights the side lobe suppression level, persistent at -13dB}
     \label{fig:QAM_rc}
\end{figure*}
\subsection{Contributions}
The main contributions can be summarized as follows:
\begin{itemize} 
    \item Show the potentiality of the ISAC approach by applying it to the context of UAV-based communication to generate SAR images, staying within the boundaries of common standards~\cite{rel18, dsrc}.
    \item Produce simulated SAR images with the QUAsi Deterministic RadIo channel GenerAtor (QuaDRiGa) \cite{jaeckel2014quadriga} 5G-standard-compliant channel model and investigate performance analyses measuring Bit Error Rate (BER) and Noise Equivalent Sigma Zero (NESZ) of the system under various conditions, furnishing a numerical assessment of the system performances. We present a scenario involving the potential burial of one or more individuals beneath the snow, as might occur after an avalanche.
    \item Provide the first ISAC implementation of a communication waveform for environmental imaging employing a UAV equipped with a standard sub-6 GHz link, utilizing norm-compliant transmitted powers, a prescribed number of antennas, and their respective directivity and gains.
    \item Present the results of an experimental campaign to validate what is proposed and show the UAV-based setup's effectiveness while emulating an Orthogonal Frequency Division Multiplexing (OFDM) transmission by convolving actual radar chirp measurement onboard a UAV. 
\end{itemize}
\subsection{Paper structure}
Section~\ref{sec:signal_modelling} delineates the waveform characterization and the parameters used. Subsequently, we present some principles of SAR imaging.  In Section~\ref{sec:performance_analysis}, we furnish a numerical assessment of the system performances in multiple scenarios involving the burial of individuals beneath the snow. Section~\ref{sec:simulations} highlights a simulator employing the standard-compliant QuaDRiGa channel generator to perform SAR focusing and study the system's Impulse Response Function (IRF). In Section~\ref{sec:experimental}, we culminate the paper by showcasing actual data acquired by a UAV equipped with a Software-Defined Radar (SDR) payload.
Finally, Section~\ref{sec:conclusions} summarizes the results and provides final remarks.  

\section{Signal Modeling and SAR imaging}
\label{sec:signal_modelling}

This section comprehensively explains the chosen communication signal model, including standardized values for critical parameters. 
Furthermore, we present an evaluation of two different pulse compression techniques and the main algorithm used to generate a SAR image.

\subsection{OFDM signal model and parameters}

The foundation of most digital communication protocols is OFDM~\cite{tutorialBoban}, chosen for its resilience to wireless channel impairments and versatile design in both the time and frequency domains. 
The time domain's most general OFDM transmission structure is arranged into frames, subframes, and slots. All the details of the internal definition of the OFDM frame can be found in~\cite{tutorialBoban,ETSIphy}.
 


Regarding its physical layer parameters, which encompass aspects like bandwidth, sub-carrier spacing, symbol duration, receiver noise figure, and more, detailed information is provided in Sec.~\ref{sec:simulations}. These values are sourced from contemporary literature and established standards~\cite{tutorialBoban, dsrc, ahangar2021survey}.


To facilitate the reader, we report the definition of a single OFDM symbol as  
\begin{align}  
x(t)=\sum^{M-1}_{m=0} s(m)e^{j 2 \pi m \Delta f t} g(t)
\end{align}
where $s(m)$ is the constellation symbol (QAM, QPSK, etc.), $\Delta f = 1/T_s$ is the sub-carrier spacing defined by the inverse of the symbol interval $T_s$, $M$ is the number of subcarriers, and $g(t)$ is the pulse shaping filter. The actual transmitted waveform is generated by attaching the Cyclic Prefix (CP), with a length of 6.57\% of the OFDM symbol, as defined by the standard.

%

\begin{figure*}[!htbp]
    \centering
    \captionsetup{justification=centering}
    \begin{subfigure}[t]{\columnwidth}    
                  \centering
        \includegraphics[width=0.8\columnwidth]{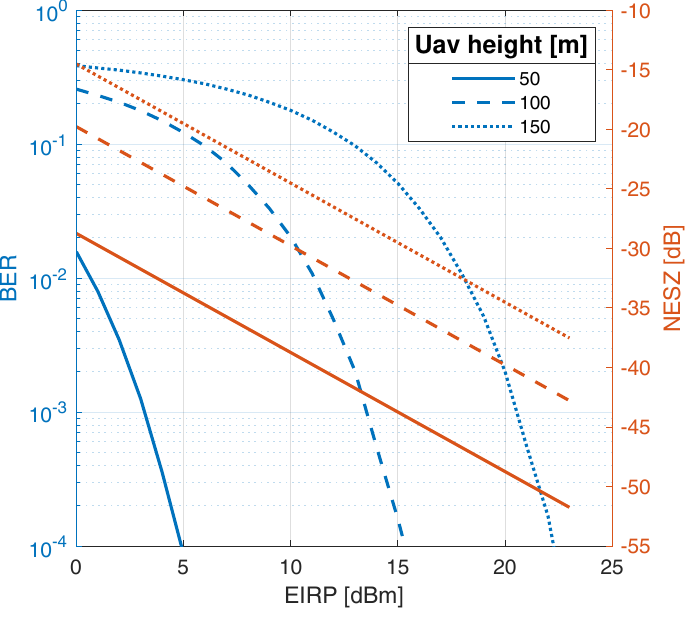}
        \caption{}
        \label{fig:BER_NESZ_altitude}
    \end{subfigure}
    \begin{subfigure}[t]{\columnwidth}
                \centering
        \includegraphics[width=0.8\columnwidth]{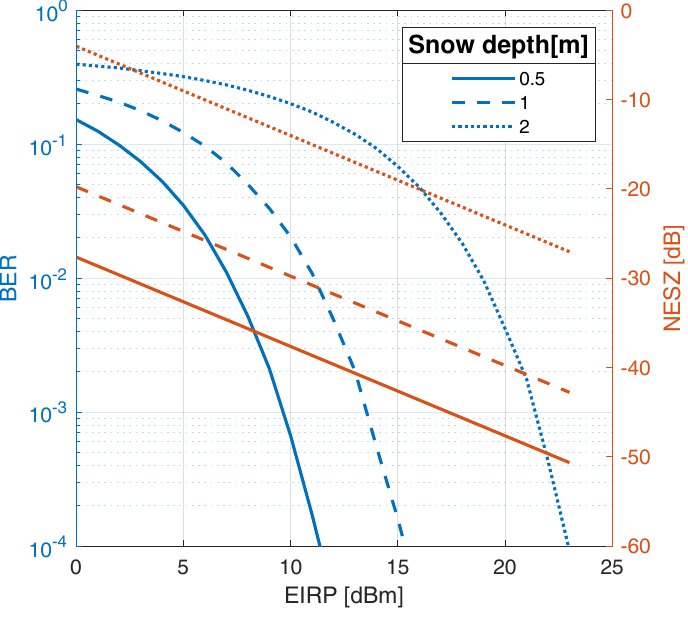}
        \caption{}
        \label{fig:BER_NESZ_snow_depth}
    \end{subfigure}
    \caption{In~(\ref{fig:BER_NESZ_altitude}), the BER and NESZ vs EIRP for different UAV heights at fixed snow depth of 1 meter. In~(\ref{fig:BER_NESZ_snow_depth}), the BER and NESZ vs EIRP for different snow depths. The nominal flying altitude of the UAV is 100 m. The extinction ratio in the snow can be found in \cite{tiuri_complex_1984}}
\end{figure*}
\subsection{Pulse compression of OFDM signals}
\label{sec:pulse_compression}
The process chain toward obtaining a focused SAR image comprises two main steps: the range and the azimuth compression. The former is commonly called \textit{matched filtering} in communication theory. Several studies are present in the literature addressing this topic for OFDM signals~\cite{rodriguez_supervised_2023, rodriguez_experimental_2023}. 
We present here two possible range compression methods. The first employs the conventional matched filter, while the second is Zero Forcing (ZF), a well-known approach in digital communication that consists of an inversion of the frequency response of the pulse.

\subsubsection{Matched filter}
For the matched filter case, the received signal is multiplied in the frequency domain by the complex conjugate of the transmitted one
\begin{equation}
    X_{\text{RC}}(f) = X^*(f)X_{\text{RX}}(f),
\end{equation}
where $X(f)$ and $X_{\text{RX}}(f)$ represent the frequency domain representations of the transmitted and received signal, respectively. The inverse Fourier Transform of $X_{\text{RC}}(f)$ leads to the range compressed signal. With this procedure, the lack of spectrum equalization can lead to possible performance degradation. With perfect equalization, we expect to have an IRF, also defined as the auto-correlation of the pulse, that closely resembles the ideal cardinal sine function. However, when each subcarrier has a different channel gain, we have an unbalanced power spectrum. This will generate higher sidelobes in the IRF that cause interference between targets close in range. All this translates to a decreased quality in the generated SAR image. 

\subsubsection{Zero Forcing}
One of the alternatives commonly used is the  ZF receiver. With this approach, we multiply the received signal by the inverse of the reference waveform
\begin{equation}
    X_{\text{RC}}(f) = \frac{1}{X(f)}X_{\text{RX}}(f).
\end{equation}
One possible drawback is the presence of very high gain values in the inverse filter due to a very small amplitude in the reference signal. To mitigate this, we can rewrite the inverse procedure as
\begin{equation}
        X_{\text{RC}}(f) = \frac{1}{X(f)+k}X_{\text{RX}}(f),
\end{equation}
where $k$ is computed based on the SNR and the channel response~\cite{mark2003weihua}. 
A simulation has been carried out to compare these two approaches, and the results using an OFDM waveform composed of 1024 sub-carriers are depicted in Fig.~\ref{fig:QAM_rc}.
Figure~\ref{fig:QPSK_corr} reports the comparison when symbols $s(m)$ are drawn from a QPSK constellation. We can see that the ZF filter provides a better side-lobes rejection than the matched filter, even in the case of a quasi-flat spectrum, leading to an almost perfect sinc-like IRF.

In Fig.~\ref{fig:QAM_corr}, we present instances of the pulse-compressed signal employing 256-QAM modulated sub-carriers. In this case, the noise floor after matched filtering is higher than in the previous example due to power imbalances between subcarriers. The ZF, on the other hand, leads once again to the desired IRF without sidelobes and a much lower noise floor.

\subsection{SAR Image Formation}
The second step of SAR imaging is called \textit{SAR focusing}. We can employ the Time Domain Back Projection (TDBP) algorithm, which is suitable even for non-linear trajectories, like those typical of lightweight UAV systems. In TDBP, every pulse (range-compressed OFDM symbol) is interpolated on a grid that represents the area of interest and then re-phased, utilizing as a prior the platform's trajectory. All interpolated and re-phased pulses are then coherently summed, leading to the final high-resolution SAR image.\\
For a specified pixel at position $x,y$, we can define the TDBP as:
\begin{equation}
    F(x,y) = \int x_{\text{RC}}\left(t=\frac{R(\tau:x,y)}{c}, \tau\right) e^{j\frac{4\pi}{\lambda}R(\tau:x,y)} d\tau
\end{equation}
where $F(x,y)$ is the 2D SAR image, $c$ the speed of light, $\lambda$ the wavelength and $R(\tau:x,y)$ the nominal radar-to-pixel distance at instant $\tau$ for the $x,y$ pixel. The range-compressed signal is $x_{\text{RC}}(t, \tau)$. Notice that $t$ is the time instant within a single pulse (also called \textit{fast-time} in radar jargon), while $\tau$ is the time axis sampled every PRI (also called \textit{slow-time}). \\
The primary downside of this algorithm lies in its computational complexity. For each pulse, it is necessary to compute the radar-to-pixel distances and then make an interpolation followed by a complex multiplication. Assuming a grid with dimensions of $N\times N$ pixels and an aperture of $N$ positions, the number of operations is roughly $N^3$. More efficient algorithms, such as the Fast Factorized Back Projection introduced in~\cite{ulander2003synthetic}, can produce equivalent images with fewer computations.

\section{Performance analysis}
\label{sec:performance_analysis}
One of the most used measurements of the sensitivity of a radar system is the so-called NESZ. It corresponds to the equivalent normalized radar cross section ($\sigma_0$) that yields a unitary Signal-to-Noise Ratio (SNR) of the resulting image~\cite{miranda2016sentinel}. In any radar system, the received power can be modeled as:
\begin{equation}
    P_{\rm{RX}} = \frac{P_{\rm{TX}} G_{\rm{TX}} f(\theta)}{4 \pi R^2} \sigma \frac{A_e f(\theta)}{4 \pi R^2}
\end{equation}
where $P_{\rm{TX}}$ is the transmitted power, $G_{\rm{TX}}$ is the antenna gain of the transmitter, $f(\theta)$ is the antenna directivity function, $\sigma$ is the target's Radar Cross Section (RCS), $A_e$ is the equivalent antenna area of the receiver and $R$ the radar-to-target distance.
As known from signal theory, after range compression (or matched filtering), the signal power increases by a factor $T_p^2$, while the noise power is equal to $P_n = N_0T_p$, leading to a \rm{SNR} equal to:
\begin{equation}
    \rm{SNR}_{\text{RC}} = \frac{P_{\rm{RX}} T_p}{N_0}
\end{equation}
where $T_p$ is the pulse length, and $N_0$ is the noise power spectral density. Similarly, after azimuth compression, we have the final focused image where the $\rm{SNR}$ is increased once again by the number of processed pulses $N_\tau$
\begin{equation}
    \rm{SNR}_{\text{foc}} = \rm{SNR}_{\text{RC}} N_\tau
\end{equation}
Finally, we can get the $\rm{NESZ}$ as:
\begin{equation}
        \rm{NESZ} =     \frac{\sigma_0}{\rm{SNR_{\text{foc}}}} .
\end{equation}
where $\sigma_0 = \sigma/(\rho_{\rm{rg}} \rho_{\rm{az}})$, in which $\rho_{\rm{rg}}$ and $\rho_{\rm{az}}$ are the range and azimuth resolutions respectively. 
We assessed the imaging capability of the aerial base station by computing the NESZ, utilizing parameters that align with real-world implementations of communication standards \cite{tutorialBoban, 3GPPNR, dsrc}. As reported in Table~\ref{tab:parameters for the evaluation}, the Equivalent Isotropic Radiated Power (EIRP) ranges from 0 to the maximum prescribed limit of 23 dBm.

\begin{figure}[tb]
     \centering
     \captionsetup{justification=centering}
     \begin{subfigure}[b]{0.49\columnwidth}
         \centering
         \includegraphics[width=\textwidth]{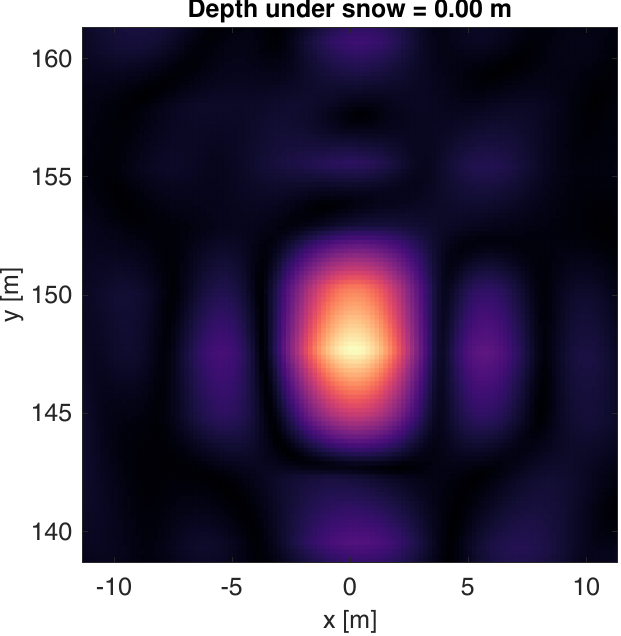}
         \caption{}
         \label{fig:SAR_noise_0}
     \end{subfigure}
\hfill
     \begin{subfigure}[b]{0.49\columnwidth}
         \centering 
         \includegraphics[width=\textwidth]{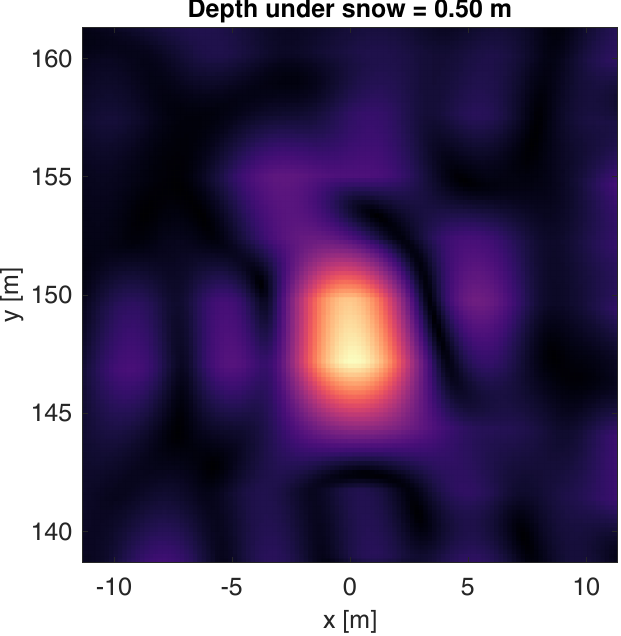}
         \caption{}
         \label{fig:SAR_noise_1}
     \end{subfigure}
     \par\medskip
          \begin{subfigure}[b]{0.49\columnwidth}
         \centering
         \includegraphics[width=\textwidth]{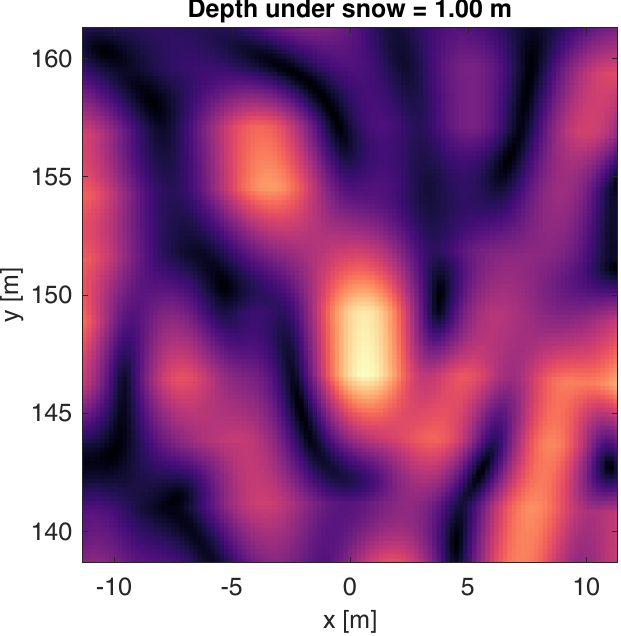}
         \caption{}
         \label{fig:SAR_noise_15}
     \end{subfigure}
     \hfill
          \begin{subfigure}[b]{0.49\columnwidth}
         \centering
         \includegraphics[width=\textwidth]{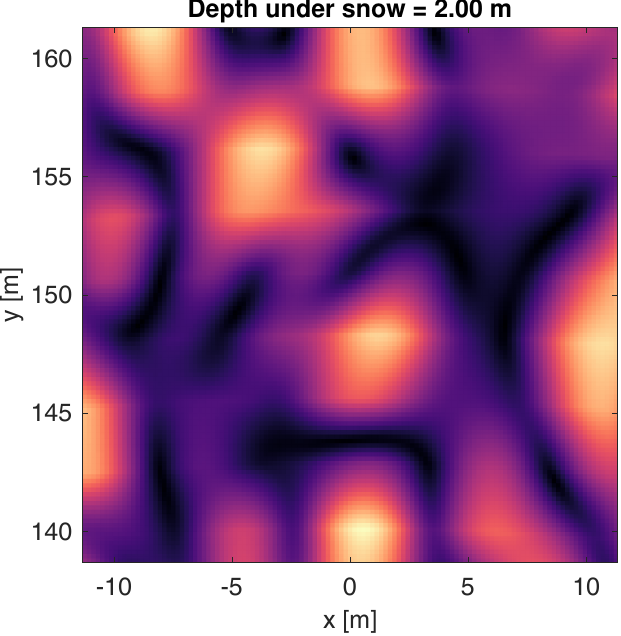}
         \caption{}
         \label{fig:SAR_noise_2}
     \end{subfigure}
     \caption{Simulated SAR images for different target depths under the snow at a nominal UAV flying altitude of 150 m.}
     \label{fig:SAR_noise}
\end{figure}
We used the widely known QuaDRiGa channel generator to accurately replicate the specific scenario of interest~\cite{jaeckel2014quadriga}. QuaDRiGa generates channel coefficients and their corresponding delays, employing a "statistical ray-tracing model". Through this approach, we achieve three-dimensional propagation, geometric polarization, continuous temporal evolution, and spatially correlated slow and fast fading. 

Two pivotal parameters for computing the NESZ involve the system's symbol duration and Pulse Repetition Frequency (PRF). We must first establish what we define as a pulse and determine its duration to attain a communication system. There are two possible approaches to consider.
The symbol-based pulse defines an equivalent radar pulse as a single OFDM symbol lasting $8~\mu$ s. This translates to a PRF of 125 kHz and a duty cycle of 100\%. The unambiguous range, defined as the longest range to which a transmitted pulse can travel out to and back again between consecutive pulses, is calculated as $R_{\text{amb}} = c/(2~\text{PRF}) = 1.2$~km.
The second approach, the frame-based pulse, takes an entire frame consisting of $N$ OFDM symbols as the equivalent radar pulse. This increases the pulse length by a factor of $N$ and consequently extends the unambiguous range. Additionally, a frame-based pulse offers greater flexibility by allowing the transmitter to incorporate non-transmission intervals between frames, which is not viable in a symbol-based pulse.
We can easily note that both approaches yield the same NESZ. In the symbol-based approach, the symbol duration is shorter, thus resulting in a slight SNR gain after range compression. However, the PRF increases, leading to more pulses integrated with azimuth compression. In contrast, the frame-based approach exhibits the opposite: the single pulse is $N$ times longer, resulting in a much higher gain after pulse compression but with a lower PRF. Consequently, the SNR and the NESZ remain equivalent for both approaches.
\begin{table}[b!]
\caption{Communication standard parameters}
\label{tab:parameters for the evaluation}
\centering
\resizebox{\linewidth}{!}{%
    \begin{tabular}{cc}
    \textbf{Parameter} & \textbf{Value}\\ 
    \hline
        Bandwidth $B$          & 40 MHz    \\
        $M_{\rm FFT}$  &  64 \\
        Subcarrier spacing $\Delta f$  & $120$ KHz     \\
        Number of subcarrier $M$            & 52  \\
         Data symbol duration   $T$     & 8.33 $\mu$s    \\
         $M_{\rm CP}$  & 12 \\
          Noise Figure & 7 dB  \\
          $f_c$ & 5.9 GHz  \\
         $EIRP_{max}$ & 23 dBm \\
         $G_{\rm tmax}$ & 10 dB \\
         QuaDRiGa NTN scenario & QuaDRiGa\_NTN\_Rural\_NLOS \\
         QuaDRiGa standard scenario  & 5G-ALLSTAR\_Rural\_LOS\\
    \hline
    \end{tabular}%
}
\end{table}

\begin{figure*}[tb]
    \centering
    \captionsetup{justification=centering}
    \begin{subfigure}[b]{0.6\columnwidth}    
        \centering
        \includegraphics[width=\columnwidth]{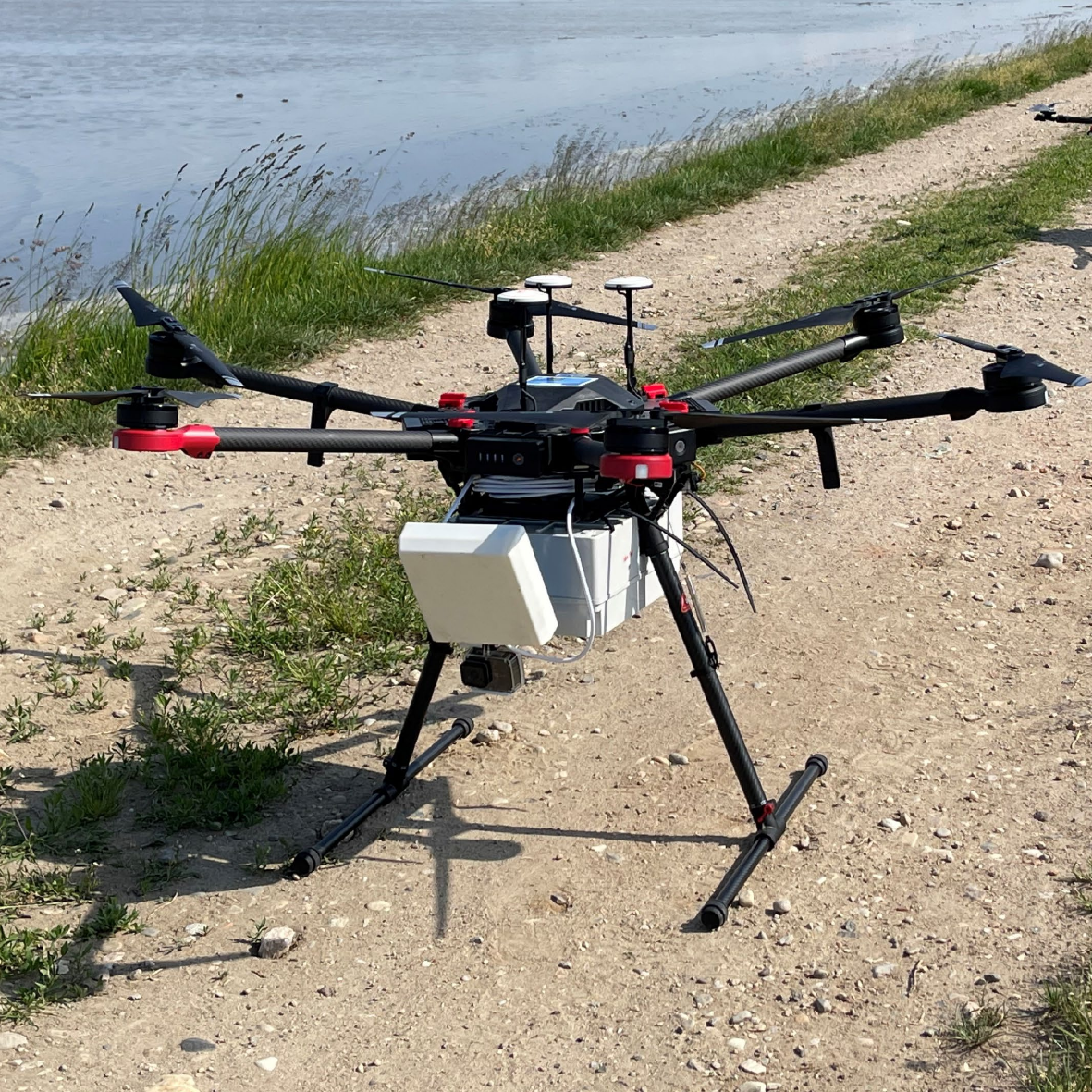}
        \caption{}
        \label{fig:experimental_setup}
    \end{subfigure}
    \hfill
    \begin{subfigure}[b]{0.6\columnwidth}
    \centering
    \includegraphics[width=\columnwidth]{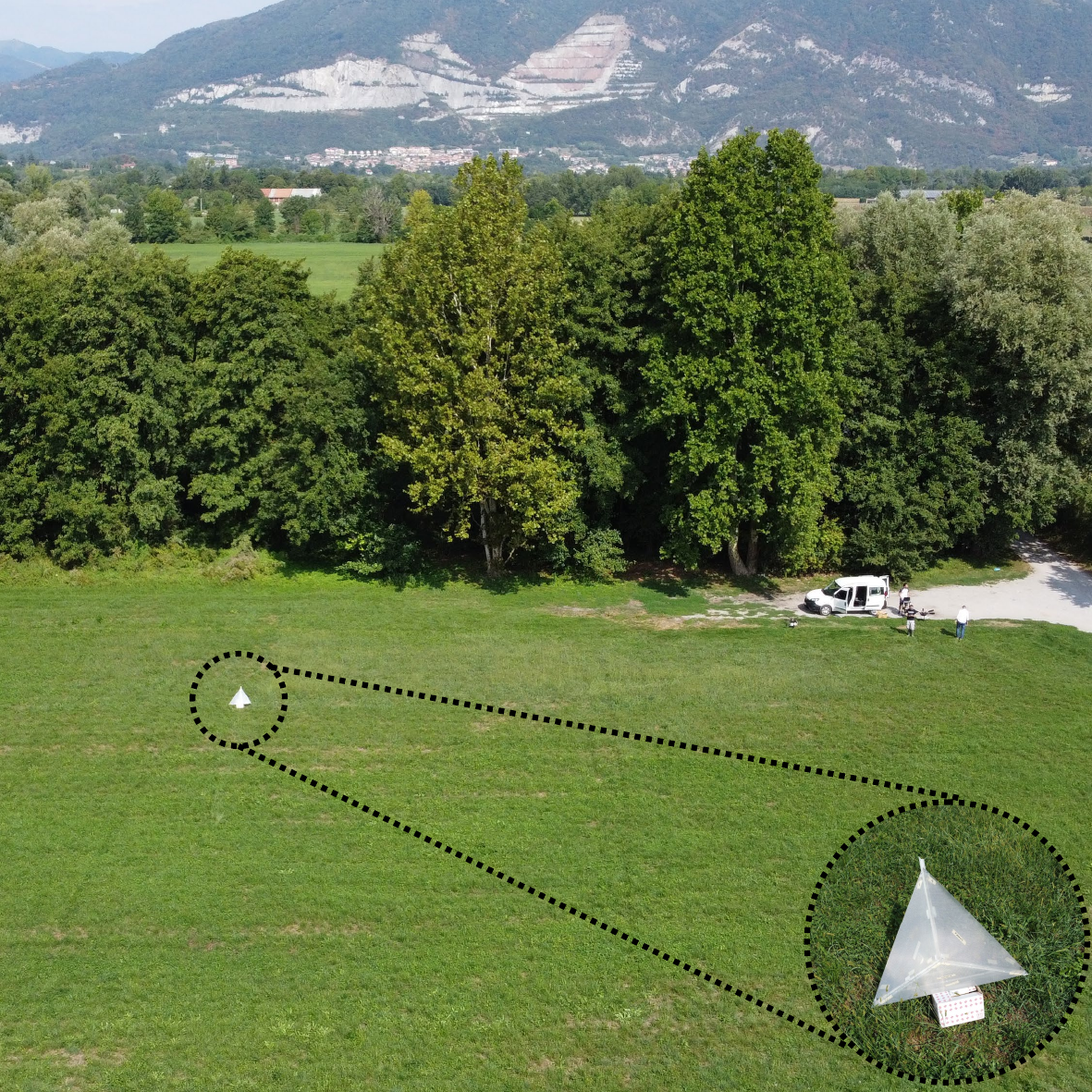}
    
    \caption{}
    \label{fig:image_scene}
    \end{subfigure}
\hfill
     \begin{subfigure}[b]{0.6\columnwidth}
    \centering
    \includegraphics[width=\columnwidth]{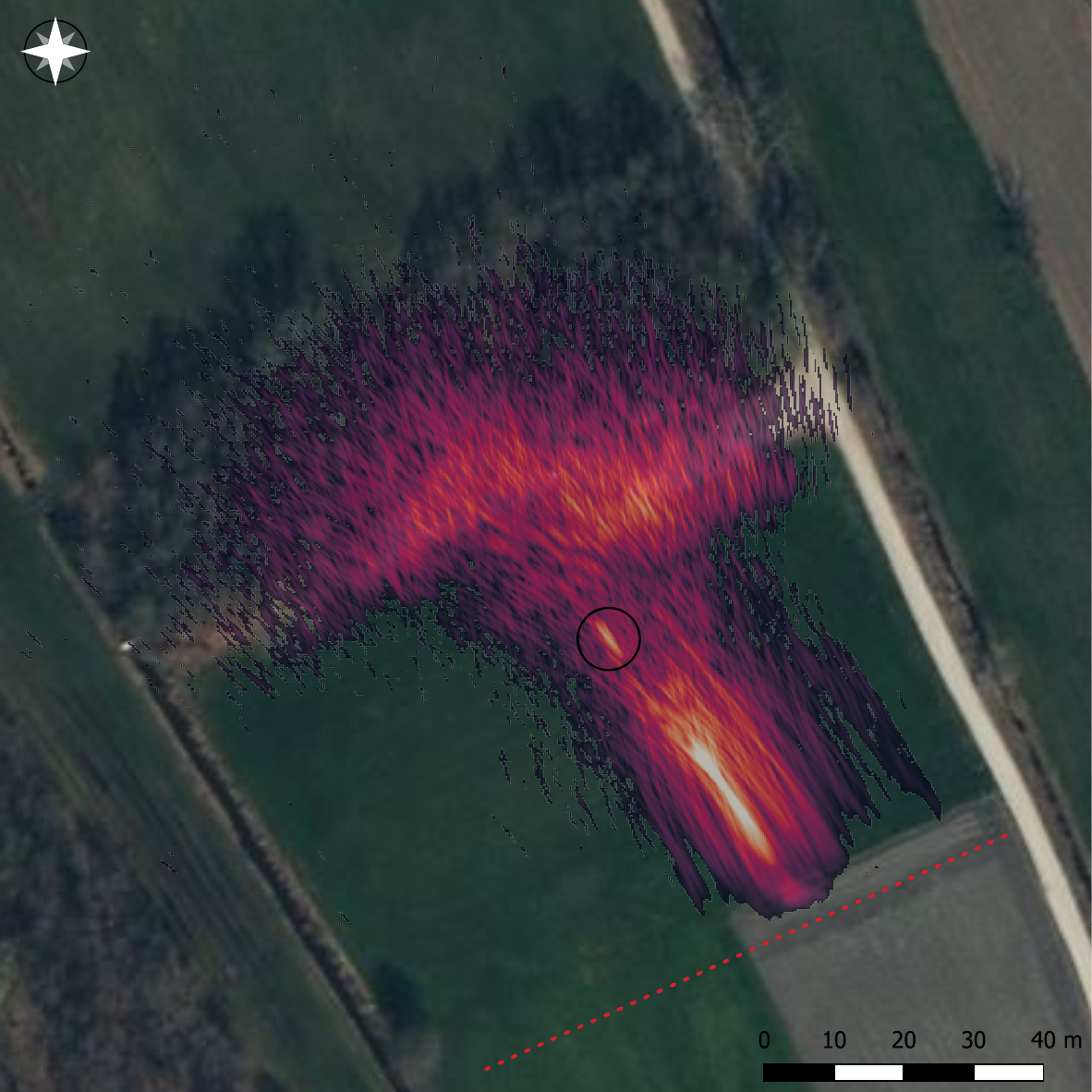}
         \caption{}
     \label{fig:radar_map}
     \end{subfigure} 
     \caption{In~(\ref{fig:experimental_setup}), the UAV is equipped with the SDR board enclosed underneath it. In~(\ref{fig:image_scene}), the optical image of the scene acquired by a camera onboard the flying UAV with a detail of the S-band Corner Reflector.  In~(\ref{fig:radar_map}), a radar image generated using TDBP processing of the OFDM signal superimposed on an optical satellite view. The red dotted line indicates the drone's trajectory and the black circle highlights a corner reflector.}
\end{figure*}

Our simulation also considered the air-to-ground communication link's Bit Error Rate (BER). Figure~\ref{fig:BER_NESZ_altitude} evaluates both metrics (NESZ and BER) for varying EIRP and UAV nominal heights. The graph's straight line indicates the NESZ shows a linear dependency on the EIRP, while the nominal height introduces a consistent shift to the NESZ curves, resulting in lower received power for longer distances.

Furthermore, we aimed to assess the system's performance in more challenging environments. One such scenario, where a UAV-based ISAC system would prove invaluable, involves situations where one or more individuals are trapped beneath the snow, for example, after an avalanche. In this context, the base station on the UAV can establish a crucial communication link for the search and rescue team while simultaneously conducting sensing operations beneath the snow. In this scenario, the BER measurement proves the feasibility of establishing a communication link with a buried user.

Figure~\ref{fig:BER_NESZ_snow_depth} illustrates the NESZ and BER evaluation with the drone flying 100 meters above the scene, varying the depth of the trapped individuals from 0.5 to 2 m. The power attenuation experienced by the signal traversing through dry snow is calculated according to the model outlined in \cite{tiuri_complex_1984}, and it is added to the propagation attenuation provided by the QuaDRiGa channel simulator. All other parameters remain consistent with those in Fig.~\ref{fig:BER_NESZ_altitude}. Our results show that subtle targets can be detected for shallow snow depths and reasonable transmitted power. However, the performance degrades rapidly as the snow depth approaches 2\,m. 

\section{Simulation results}
\label{sec:simulations}
This section presents the outcomes of a simulation conducted with the same parameters utilized in the NESZ calculation detailed in Sec.~\ref{sec:performance_analysis} and listed in Table~\ref{tab:parameters for the evaluation}. To emulate a realistic scenario, we set the transmitted power within the allowable limit of the standard, and the receiver noise figure to a value of a Commerical-Off-The-Shelf (COTS) hardware.
We employed the QPSK modulation scheme within this simulator to generate OFDM symbols. Our simulated scenarios were constructed, featuring a point scatterer set at an off-nadir angle of 45$^\circ$. This target emulates the presence of a buried person, and in ideal conditions, we expect to generate an image representing a perfect bidimensional cardinal sine. It is characterized by a Radar Cross Section (RCS) of $\sigma = 1 ~\text{m}^2$. Depending on the specific simulation, we positioned the target on the terrain's surface or buried beneath the snow. This variability in target placement allowed us to explore the system's responses to diverse environmental conditions, with target depths spanning from 0.5 to 2 meters.

In Figure~\ref{fig:SAR_noise_0}, we can see a target sharply focused, with the IRF of the system when the target is located out of the snow. As anticipated, the IRF recalls the bi-dimensional cardinal sine function we would expect by a simple point scatterer, which is typical of a SAR imaging system. In Figures~\ref{fig:SAR_noise_1}, \ref{fig:SAR_noise_15}, and \ref{fig:SAR_noise_2}, the target is submerged beneath 0.5, 1, and 2 m of snow, respectively. It is evident that the ticker the level of snow present between the target and the drone, the more the SNR and NESZ degrade. This, of course, makes target detection a challenging activity. 

\section{Experimental activity results}
\label{sec:experimental}
In this section, we present the outcomes of our experimental campaign. We established a configuration where a stationary transmitter, mounted on a 4 m pole, illuminated an area, and a receiver was fixed to a flying drone at an altitude of 40 meters. In radar jargon, this setup is defined as bistatic because the transmitter and receiver are not co-located. Both transceivers were equipped with the USRP B210 Software Defined Radio (SDR), each complemented by a 6 dBi patch antenna. The UAV employed for this survey was the DJI Matrice 600 Pro, visible in Fig.~\ref{fig:experimental_setup}, and the scene of interest is depicted in Fig.~\ref{fig:image_scene}.

During this measurement campaign, the transmitter sends a radar waveform, employing the standard chirp function defined as
\begin{equation}
    g^{\text{chirp}}(t) = \mathrm{rect} \left(\frac{t}{T_p}\right)e^{j \pi \alpha t^2},
\end{equation}
where $\alpha = B/T_p$, $B$ the total bandwidth, $T_p$ the pulse length, and the receiver collects the scene's reflections. Even in this setup, we can emulate an OFDM transmission starting from the chirp data. For each echo, we must convolve the OFDM pulse with the chirp-based range compressed signal as
\begin{equation}
    x_{RC}^{\text{OFDM}}(t) = x_{RC}^{\text{chirp}}(t) * g^{\text{OFDM}}(t) . 
\end{equation}
With this approach, we are employing the range-compressed data matrix as a continuous channel definition of our scene. Afterward, we can range-compress the signal again with a matched or ZF filter as detailed in Sec.~\ref{sec:pulse_compression}. An important step is to adapt the PRF of the SDR to the communication's protocol specifications. Finally, we can generate the image shown in Fig.~\ref{fig:radar_map} using the TDBP algorithm. As expected, it appears as a line in the focused SAR image because all those pixels have the same bistatic range. The encircled target is the corner reflector installed on the field, as shown in Fig.~\ref{fig:image_scene}. Additionally, we observe the reflection of a small woodland area nearby.

\section{Conclusion}
\label{sec:conclusions}
This research bridges digital communication and sensing, opening avenues for innovative applications by delving into the imaging capabilities of a UAV-based ISAC system. The investigation starts with defining the transmitted OFDM signal and providing realistic, standard-compliant values for crucial parameters like transmitted power, bandwidth, PRF, number of sub-carriers, and antenna gains. Two distinct pulse compression methods for OFDM signals are introduced: the traditional matched filtering and the zero forcing technique. The paper also thoroughly analyzes the performances of the setup, studying the NESZ and BER for a base station operating under challenging conditions. This assessment is carried out by varying the UAV flying altitudes and the snow depth. The results yield promising results with NESZ and BER  below -20 dB and $10^{-4}$, respectively, affirming the system's capability to detect and communicate with endangered users.
Finally, the study provides compelling evidence from simulated and experimental data affirming the UAV-based ISAC system's capability to image its environment while accurately adhering to established industry standards.
\section*{Acknowledgment}
We are glad to acknowledge that the European Union partially supported this work under the Italian National Recovery and Resilience Plan (NRRP) of NextGenerationEU, partnership on “Telecommunications of the Future” (PE00000001 - program “RESTART”) CUP: D43C22003080001, Structural Project S 13 ISaCAGE.

\bibliographystyle{IEEEtran}
{\footnotesize
\bibliography{ISAC_UAS}}

\begin{thebibliography}{10}
\providecommand{\url}[1]{#1}
\csname url@samestyle\endcsname
\providecommand{\newblock}{\relax}
\providecommand{\bibinfo}[2]{#2}
\providecommand{\BIBentrySTDinterwordspacing}{\spaceskip=0pt\relax}
\providecommand{\BIBentryALTinterwordstretchfactor}{4}
\providecommand{\BIBentryALTinterwordspacing}{\spaceskip=\fontdimen2\font plus
\BIBentryALTinterwordstretchfactor\fontdimen3\font minus
  \fontdimen4\font\relax}
\providecommand{\BIBforeignlanguage}[2]{{%
\expandafter\ifx\csname l@#1\endcsname\relax
\typeout{** WARNING: IEEEtran.bst: No hyphenation pattern has been}%
\typeout{** loaded for the language `#1'. Using the pattern for}%
\typeout{** the default language instead.}%
\else
\language=\csname l@#1\endcsname
\fi
#2}}
\providecommand{\BIBdecl}{\relax}
\BIBdecl

\bibitem{meng_uav-enabled_2023}
K.~Meng, Q.~Wu, J.~Xu, W.~Chen, Z.~Feng, R.~Schober, and A.~L. Swindlehurst,
  ``{UAV}-enabled integrated sensing and communication: Opportunities and
  challenges,'' pp. 1--9, conference Name: {IEEE} Wireless Communications.

\bibitem{petritoli_reliability_2017}
E.~Petritoli, F.~Leccese, and L.~Ciani, ``Reliability assessment of {UAV}
  systems,'' in \emph{2017 {IEEE} {International} {Workshop} on {Metrology} for
  {AeroSpace} ({MetroAeroSpace})}.\hskip 1em plus 0.5em minus 0.4em\relax
  Padua, Italy: IEEE, Jun. 2017, pp. 266--270.

\bibitem{zeng_wireless_2016}
Y.~Zeng, R.~Zhang, and T.~J. Lim, ``Wireless communications with unmanned
  aerial vehicles: opportunities and challenges,'' vol.~54, no.~5, pp. 36--42,
  conference Name: {IEEE} Communications Magazine.

\bibitem{trotta2017fly}
A.~Trotta, M.~Di~Felice, K.~R. Chowdhury, and L.~Bononi, ``Fly and recharge:
  Achieving persistent coverage using small unmanned aerial vehicles (suavs),''
  in \emph{2017 IEEE ICC}.\hskip 1em plus 0.5em minus 0.4em\relax IEEE, 2017,
  pp. 1--7.

\bibitem{cui_integrating_2021}
Y.~Cui, F.~Liu, X.~Jing, and J.~Mu, ``Integrating sensing and communications
  for ubiquitous {IoT}: Applications, trends, and challenges,'' vol.~35, no.~5,
  pp. 158--167, conference Name: {IEEE} Network.

\bibitem{grathwohl_taking_2022}
A.~Grathwohl, M.~Stelzig, J.~Kanz, P.~Fenske, A.~Benedikter, C.~Knill,
  I.~Ullmann, I.~Hajnsek, A.~Moreira, G.~Krieger, M.~Vossiek, and
  C.~Waldschmidt, ``Taking a look beneath the surface: Multicopter {UAV}-based
  ground-penetrating imaging radars,'' vol.~23, no.~10, pp. 32--46, conference
  Name: {IEEE} Microwave Magazine.

\bibitem{linsalataUAV}
F.~Linsalata, A.~Albanese, V.~Sciancalepore, F.~Roveda, M.~Magarini, and
  X.~Costa-Perez, ``Otfs-superimposed prach-aided localization for uav safety
  applications,'' in \emph{2021 IEEE GLOBECOM}, 2021, pp. 1--6.

\bibitem{pin_tan_integrated_2021}
D.~K. Pin~Tan, J.~He, Y.~Li, A.~Bayesteh, Y.~Chen, P.~Zhu, and W.~Tong,
  ``Integrated sensing and communication in 6g: Motivations, use cases,
  requirements, challenges and future directions,'' in \emph{2021 1st {IEEE}
  International Online Symposium on {JC}\&S}, pp. 1--6.

\bibitem{rel18}
\BIBentryALTinterwordspacing
3rd Generation Partnership~Project. (2022) {3GPP Release 18}. [Online].
  Available: \url{https://www.3gpp.org/release18}
\BIBentrySTDinterwordspacing

\bibitem{dsrc}
J.~B. Kenney, ``Dedicated short-range communications (dsrc) standards in the
  united states,'' \emph{Proceedings of the IEEE}, vol.~99, no.~7, pp.
  1162--1182, 2011.

\bibitem{jaeckel2014quadriga}
S.~Jaeckel, L.~Raschkowski, K.~B{\"o}rner, and L.~Thiele, ``Quadriga: A 3-d
  multi-cell channel model with time evolution for enabling virtual field
  trials,'' \emph{IEEE transactions on antennas and propagation}, vol.~62,
  no.~6, pp. 3242--3256, 2014.

\bibitem{tutorialBoban}
M.~H.~C. Garcia, A.~Molina-Galan, M.~Boban, J.~Gozalvez, B.~Coll-Perales,
  T.~Şahin, and A.~Kousaridas, ``A tutorial on {5G NR V2X} communications,''
  \emph{IEEE Communications Surveys \& Tutorials}, vol.~23, no.~3, pp.
  1972--2026, 2021.

\bibitem{ETSIphy}
3GPP, ``{NR}; physical channels and modulation (release 15),'' 3rd Generation
  Partnership Project {(3GPP)}, Technical Specification {(TS)} 38.211, Jan.
  2020, version 15.8.0.

\bibitem{ahangar2021survey}
M.~N. Ahangar, Q.~Z. Ahmed, F.~A. Khan, and M.~Hafeez, ``A survey of autonomous
  vehicles: Enabling communication technologies and challenges,''
  \emph{Sensors}, vol.~21, no.~3, p. 706, 2021.

\bibitem{tiuri_complex_1984}
M.~Tiuri, A.~Sihvola, E.~Nyfors, and M.~Hallikaiken,
  ``\BIBforeignlanguage{en}{The complex dielectric constant of snow at
  microwave frequencies},'' \emph{\BIBforeignlanguage{en}{IEEE Journal of
  Oceanic Engineering}}, vol.~9, no.~5, pp. 377--382, Dec. 1984.

\bibitem{rodriguez_supervised_2023}
J.~T. Rodriguez, F.~Colone, and P.~Lombardo, ``Supervised {Reciprocal} {Filter}
  for {OFDM} {Radar} {Signal} {Processing},'' \emph{IEEE Transactions on
  Aerospace and Electronic Systems}, vol.~59, no.~4, pp. 3871--3889, Aug. 2023.

\bibitem{rodriguez_experimental_2023}
------, ``Experimental evaluation of {Supervised Reciprocal Filter Strategies
  for OFDM-radar} signal processing,'' in \emph{2023 IEEE {RadarConf23}}.\hskip
  1em plus 0.5em minus 0.4em\relax IEEE, 2023, pp. 1--6.

\bibitem{mark2003weihua}
J.~Mark and W.~Zhuang, ``Wireless communications and networking".''\hskip 1em
  plus 0.5em minus 0.4em\relax Prentice Hall, 2003.

\bibitem{ulander2003synthetic}
L.~M. Ulander, H.~Hellsten, and G.~Stenstrom, ``Synthetic-aperture radar
  processing using fast factorized back-projection,'' \emph{IEEE Transactions
  on Aerospace and electronic systems}, vol.~39, no.~3, pp. 760--776, 2003.

\bibitem{miranda2016sentinel}
N.~Miranda, P.~J. Meadows, A.~Pilgrim, R.~Piantanida, A.~Recchia, D.~Giudici,
  D.~Small, and A.~Schubert, ``Sentinel-1b preliminary results obtained during
  the orbit acquisition phase [work in progress],'' \emph{Procedia Computer
  Science}, vol. 100, pp. 1313--1318, 2016.

\bibitem{3GPPNR}
{3GPP TS 38.211}, ``{NR; Physical channels and modulation},'' 2018.

\end{thebibliography}

\end{document}